%% file: quantized_cs_arxiv.tex
\documentclass[a4paper,10pt,twocolumn,twoside]{IEEEtran}
\usepackage{color}
\usepackage{graphicx}
\usepackage{psfrag}
\usepackage{algorithmic}
\usepackage[ruled]{algorithm}
\makeatletter
\def\ALG@name{Alg.}

\definecolor{commentgray}{rgb}{.3,.3,.3}

\def\algorithmiccomment#1{\hfill\llap{\hbox to 33mm{%
			\color{commentgray}// #1 \hss}}}

\makeatother
\graphicspath{{./}}
\DeclareSymbolFont{AMSb}{U}{msb}{m}{n}
\DeclareFontFamily{U}{msb}{}%
\DeclareFontShape{U}{msb}{m}{n}{<-6>msbm5<6-8>msbm7<8->msbm10}{}%
\DeclareMathSymbol{\C}{\mathalpha}{AMSb}{"43}
\DeclareMathSymbol{\N}{\mathalpha}{AMSb}{"4E}
\DeclareMathSymbol{\R}{\mathalpha}{AMSb}{"52}
\DeclareMathSymbol{\Z}{\mathalpha}{AMSb}{"5A}
\DeclareMathSymbol{\Q}{\mathalpha}{AMSb}{"51}
\def\ul#1{\smash{\underline{\vrule width 0pt height 0pt depth .1pt%
                \smash{\hbox{#1}}}}}
\def\ve#1{{\mathchoice{\mbox{\boldmath$\displaystyle #1$}}%
              {\mbox{\boldmath$\textstyle #1$}}%
              {\mbox{\boldmath$\scriptstyle #1$}}%
              {\mbox{\boldmath$\scriptscriptstyle #1$}}}}
\def\dB{\mathrm{\, dB}}
\def\T{\mathsf{T}}
\def\argmax{\mathop{\mathrm{argmax}}}
\def\argmin{\mathop{\mathrm{argmin}}}
\def\A{\mathcal{C}}
\def\AO{\mathcal{C}_0}
\def\sigmaN{\sigma_n}
\def\sigman2{1/\sigma_n^2}
\def\quant{\mathcal{Q}}
\def\S{\mathcal{S}}
\def\equiv{\mathrel{\widehat{=}}}
\def\defeq{\stackrel{\mbox{\tiny def}}{=}}
\begin{document}
\title{Discrete Sparse Signals: Compressed Sensing by Combining
		OMP and the Sphere Decoder}
\IEEEoverridecommandlockouts
\author{%
\IEEEauthorblockN{Susanne Sparrer, Robert F.H. Fischer}\\
\IEEEauthorblockA{Institute of Communications Engineering,
		Ulm University, Ulm, Germany
\thanks{Email: susanne.sparrer@uni-ulm.de, robert.fischer@uni-ulm.de}
}\vspace*{-6mm}
}
\maketitle

\begin{abstract}
We study the reconstruction of discrete-valued sparse signals from
underdetermined systems of linear equations. On the one hand, classical
compressed sensing (CS) is designed to deal with real-valued sparse signals.
On the other hand, algorithms known from MIMO communications, especially the
sphere decoder (SD), are capable to reconstruct discrete-valued non-sparse
signals from well- or overdefined system of linear equations. Hence, a
combination of both approaches is required. We discuss strategies to include
the knowledge of the discrete nature of the signal in the reconstruction
process. For brevity, the exposition is done for combining the orthogonal
matching pursuit (OMP) with the SD; design guidelines are derived. It is
shown that by suitably combining OMP and SD an efficient low-complexity
scheme for the detection of discrete sparse signals is obtained.
\end{abstract}

\section{Introduction}
\label{sec_1}

\noindent
In some applications like multiple-access schemes with a very small number of
active users (e.g., sensor networks) \cite{Zhu:2011,Schepker:2011,Knoop:2012}
or peak-to-average power ratio reduction in orthogonal frequency-division
multiplexing (OFDM) \cite{Fischer:2012} a \emph{discrete-valued} sparse signal
has to be estimated based on an under-determined system of linear equations.
Even in source coding, the direct estimation of the quantized transform-domain
coefficients may be beneficial.

Usually, in compressed sensing (CS) a \emph{real-valued} $s$-sparse (column)
vector $\ve{x} \in \R^L$ has to be reconstructed from an under-determined
system of linear equations \cite{CS}.
Specifically, if $\ve{A} \in \R^{K \times L}$ is the measurement matrix and
$\ve{y} = \ve{A}\ve{x} + \ve{n} \in \R^K$, $K \ll L$, is the noisy observation
(AWGN with variance $\sigma_n^2$ per component), the following problem has
to be solved%
\footnote{Notation: $||\cdot||_p$ denotes the $\ell_p$ norm.
	% (even though for $p = 0$ it is not even a quasinorm).
	$\ve{A}_\S$ is the matrix composed of the columns of $\ve{A}$, whose
	indices are in the set $\S$, and $\ve{x}_\S$ is the vector with the
	elements of $\ve{x}$, whose indices are in the set $\S$.
	$\bar{\S}$ is the complement of the set $\S$ w.r.t.\ $\{1,\ldots,L\}$.
	$\ve{A}^+$ denotes the Moore-Penrose (left) pseudoinverse of
	$\ve{A}$.
	$\quant_{\A}(\cdot)$: element-wise quantization to a given alphabet
	$\A$.
}
($\epsilon$: given tolerance)
\begin{equation}						\label{eq_cs_1}
\ve{\hat{x}} = \argmin_{\ve{\tilde{x}} \in \R^L}
		\left\| \tilde{\ve{x}} \right\|_0 \;,
	\quad\mbox{with}\quad
	\| \ve{A}\tilde{\ve{x}} - \ve{y}\|_2 \le \epsilon \;.
\end{equation}
This can be (approximately) done by using one of the standard CS algorithms;
in view of the computational complexity greedy approaches like the
\emph{\ul{o}rthogonal \ul{m}atching \ul{p}ursuit (OMP)} \cite{OMP} or the
\emph{\ul{co}mpressive \ul{sa}mpling \ul{m}atching \ul{p}ursuit (CoSaMP)}
\cite{CoSaMP} are of special interest.

If the non-zero elements of the sparse vector $\ve{x}$ are chosen from a finite
set $\A$ and $\AO \defeq \A \cup \{0\}$, we have to solve (\ref{eq_cs_1}) with
trial vector $\ve{\tilde{x}} \in \AO^L$. Please note that in contrast to
``one-bit CS'', e.g., \cite{Boufounos:2008}, here still $\ve{y} \in \R^K$,
i.e., the sparse vector $\ve{x}$, not the measurement vector $\ve{y}$, is
discrete.

Stating from (\ref{eq_cs_1}), the obvious strategy is to run conventional CS
to obtain a real-valued estimate followed by quantizing the elements of the
vector to the set $\AO$. The main drawback is that the knowledge about the
discrete nature of the sparse signal is not used in the reconstruction step,
which---whenever side information is ignored---causes a loss.

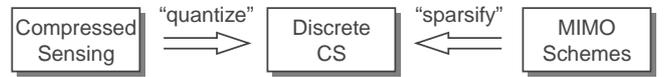
\begin{figure}[ht]
\centerline{\input{xfig_csmimo.pstex_t}}
\caption{\label{fig_csmimo}
	The connection between CS, discrete CS, and MIMO detection.
}
\end{figure}

However, the CS problem (\ref{eq_cs_1}) with trial vector
$\ve{\tilde{x}} \in \AO^L$ can be rewritten in the form
\begin{equation}						\label{eq_cs_2}
\ve{\hat{x}} = \argmin_{\ve{\tilde{x}}\in\AO^L}
		\| \ve{A}\tilde{\ve{x}} - \ve{y}\|_2\ \;,
	\quad\mbox{with}\quad
	\left\| \tilde{\ve{x}} \right\|_0 \le s \;.
\end{equation}
Looking at (\ref{eq_cs_2}), since a number of discrete signals are interfering
with each other, the field of \emph{multiple-input/multiple-output (MIMO)
schemes} has to be considered. Lattice decoding algorithms, in particular the
so-called \emph{sphere decoder (SD)} \cite{SD}, solve a well-defined or
over-determined system $\ve{y} = \ve{H}\ve{x} + \ve{n}$, where
$\ve{H} \in \R^{K \times E}$, $K \ge E$, w.r.t.\ minimum Euclidean distance.
In CS with discrete-valued signals, CS recovery and lattice decoding/MIMO
equalization meet each other. Hence, either a sparsity constraint is
introduced in the SD, or the discrete nature of the signal is incorporated
into the CS recovery algorithm, cf.\ Fig.~\ref{fig_csmimo}.

In this paper, we combine both worlds in order to benefit from the different
features they have. After briefly reviewing known approaches for adapting the
SD to sparse signals, we show how OMP (as prominent representative of greedy
recovering algorithms) can be used in connection with a SD, operating on a
much lower-dimensional problem as the known approaches. Moreover, we show that
the decoding metric in the SD has to be properly adjusted to the given
sparsity. In each case we assume that the sparsity is known, and we restrict
ourselves to binary signals, i.e., $\A = \{-1, +1\}$.

The paper is organized as follows. In Sec.~\ref{sec_2}, different methods
of incorporating quantization/SD into compressed sensing recovery algorithms
are introduced. The performance of the schemes is assessed via numerical
simulations and design guidelines are derived in Sec.~\ref{sec_3}.
Brief conclusions are drawn in Sec.~\ref{sec_4}.

\section{CS using the Sphere Decoder}
\label{sec_2}

\noindent
In this section, we first review how to directly use the SD for sparse signals
and then show how OMP can be combined with the SD to obtain a low-complexity
scheme for the detection of discrete sparse signals.

\subsection{SD with Sparsity Constraint}

In order to use the SD for sparse under-determined problems (approaching the
problem ``from the right'' in Fig.~\ref{fig_csmimo}), essentially two
modifications to the standard SD are needed. On the one hand, it has to be
adapted to the sparsity constraint. Several attempts have already been made
in the literature, e.g., \cite{Zhu:2011,Schepker:2011,Knoop:2012,Tian:2009}.
On the other hand, the standard SD does not work for under-determined systems
of linear equations. Basically, two different approaches to solve this problem
exist. First, the set of equations may artificially be enlarged to the full
dimensionality $L$, see, e.g., \cite{SD_augMatrix}. This augmentation has
to be carefully done to avoid close-to-singular matrices, which, in addition
to the huge dimensionality, dramatically increases the search complexity.

Second, the SD may only be applied to a $K$-dimensional part of $\ve{x}$,
while a brute-force search over the remaining $L-K$ components is carried out,
e.g., \cite{SDTeile,SDTeile_Verbessert}.
Unfortunately, for $L-K \gg 1$, this method also has a tremendous computational
complexity (cf.\ Section~\ref{sec_2_complexity}). A possible solution to
overcome this problem is presented subsequently.

\subsection{CS with Discrete-Value Constraint}
\label{sec_OMP_SD}

The straightforward approach to use CS for discrete-valued signals
(approaching the problem ``from the left'' in Fig.~\ref{fig_csmimo}) is to
run a conventional CS algorithm and to quantize the output to the given
alphabet $\A$ in a final step (denoted as $\quant_{\A}(\cdot)$). The procedure
is illustrated for OMP%
\footnote{To a large extent, the respective steps are also valid for CoSaMP
	and other greedy approaches.
}
in pseudocode representation in Alg.~\ref{Algo_OMP}.
\begin{algorithm}
\caption{\label{Algo_OMP}
	$\ve{\hat{x}} = \mathrm{OMP}\left(\ve{y}, \ve{A}, E\right)$
}
\begin{algorithmic}[1]
\STATE $\ve{\hat{x}} = \ve{0}$, \ $\ve{r} = \ve{y}$, \ $\S = \{\}$, \ $i=0$
		\COMMENT{init}
\WHILE{$i < E$}
	\STATE $i=i+1$
	\STATE $\tilde{\ve{x}} = \ve{A}^{\T}\ve{r}$, \
	    $\varsigma_{\mathrm{best}} =
	    \argmax_{\varsigma\in\overline{\S}} \left|\tilde{x}_{\varsigma}\right|$
	\STATE $\S = \S \cup \left\{\varsigma_{\mathrm{best}}\right\}$
		\COMMENT{extend support}
	\STATE $\ve{\hat{x}}_\S = \left(\ve{A}_\S\right)^+ \ve{y}$
		\COMMENT{estimate signal at $\S$}
	\STATE $\ve{r} = \ve{y} - \ve{A}\ve{\hat{x}}$
		\COMMENT{calculate residual}
\ENDWHILE
\STATE \label{line_quantOMP}
	$\ve{\hat{x}}_\S = \quant_{\AO}\left(\ve{\hat{x}}_\S\right)$
		\COMMENT{quantize signal}
\end{algorithmic}
\end{algorithm}

In OMP, in each iteration one new support position is added to the support
set $\S$ in a greedy fashion. Specifically, the element with the largest
correlation with the residual%
\footnote{This choice is justified from a signal representation perspective.
	When discrete symbols have to be detected, a \emph{reliability measure}
	should be used. Since for binary transmission log-likelihood ratios
	are proportional to the observation, the selection criterion is
	reasonable in the present setting.
}
is selected. Usually, knowing the sparsity, $E = s$ iterations are carried out.
One disadvantage of OMP is that, once a support element has been chosen,
it can never be removed again which leads to a degradation of the performance.
To avoid this fact, in \cite{Zhang:2011,Sparrer:2013} it has been
proposed to run some additional iterations, i.e., $E > s$, in order to be able
to find all support elements even if some wrong elements have been chosen.
Quantization has then to be done w.r.t.\ $\AO$.

\subsubsection{Obvious Concatenation}

Using this obvious concatenation of OMP and subsequent quantization (we denote
this by ``OMP/Q''), the real-valued signal estimate is the basis for the final
decisions. Please note, for each realization of the vector $\ve{\hat{x}}_\S$
the threshold of the quantizer is optimized, such that exactly $s$ non-zero
samples are obtained (fixed sparsity).

However, the estimate can be discarded and only the support set estimate $\S$
may be utilized; CS just serves for finding the support. But, knowing $\S$,
a MIMO detection problem with ``channel'' matrix
$\ve{A}_\S \in \R^{K \times |\S|}$ results; since $|\S| < K$, an over-determined
problem is present. Consequently, Line~9 may be replaced by any advanced MIMO
detection scheme (e.g., decision-feedback equalization, lattice-reduction-aided
techniques) to improve the estimate. The SD (strategy denoted by ``OMP/SD'')
is again of particular interest.

\subsubsection{Embedding the Detection}

An alternative strategy to cascading OMP and a MIMO detection scheme is to
embed the detection into the algorithms. Thereby, the knowledge about the
finite alphabet is directly utilized in the reconstruction. An obvious
procedure is an element-wise quantization of the current signal estimate
within the algorithm, i.e., to replace Line~6 of Alg.~\ref{Algo_OMP} by
$\ve{\hat{x}}_\S = \quant_{\AO}\big( \left(\ve{A}_\S\right)^+ \ve{y}\big)$
and delete Line~9 (strategy denoted by ``Q-OMP'').

In terms of communications, Line~6 is nothing else than zero-forcing (ZF)
linear equalization applied to an over-determined MIMO detection problem.
Consequently, in this step any MIMO detection strategy can be utilized.
Since channel noise is present, the minimum mean-squared error (MMSE)
approaches may be preferred over the ZF one. Advanced MIMO detection
schemes are again of special interest; in particular the SD can be employed
at this step. We denote this strategy of OMP with embedded SD by ``SD-OMP''.

\subsection{Adaptation of the Branch Metric in SD}
\label{sec_branch_metric}

Applying SD in the above settings, the branch metric should be adapted.
Specifically, the fact that the a-priori probabilities of the elements of
$\AO$ are non-uniformly distributed should be included. Applying the
maximum-a-posterior criterion gives, cf.\ \cite{Knoop:2012},
\begin{eqnarray}
\hat{\ve{x}}_\S &=&
	\argmax_{\tilde{\ve{x}}_\S\in\AO^{|\S|}} \Pr\{\tilde{\ve{x}}_\S|\ve{y}\}
	= \argmax_{\tilde{\ve{x}}_\S\in\AO^{|\S|}}
		\Pr\{\ve{y}|\tilde{\ve{x}}_\S\}\Pr\{\tilde{\ve{x}}_\S\}
\nonumber\\
&=& \argmin_{\tilde{\ve{x}}_\S\in\AO^{|\S|}} \Big\{
	\|\ve{y}-\ve{A}\tilde{\ve{x}}_\S\|_{2}^2
	- 2\sigmaN^2
	\sum\limits_{\iota=1}^{|\S|}\ln(\Pr\{\tilde{x}_{\S\left(\iota\right)}\})
	\Big\}
	\;.\quad
\end{eqnarray}
In contrast to the approach given in \cite{Knoop:2012}, where the a-priori
probability is not updated, we take already available decisions into account.
If the decoder has reached depth $\iota$ of the decoding tree, i.e., still
$j = |\S|-\iota+1$ decisions are missing and already $m$ non-zero elements
have been detected, the a-priori probability of the symbols $x_{\S(\iota)}$
is given as
\begin{equation}
\Pr\{ x_{\S(\iota)} \} = \cases{\frac{s-m}{2j}    \;, & $x_{\S(\iota)} \neq 0$ \cr
			       \frac{j-(s-m)}{j} \;, & $x_{\S(\iota)} = 0$} \;.
\end{equation}
Please note that this approach guarantees $\ve{\hat{x}}$ to have the desired
(known) sparsity $s$.

\subsection{Complexity Analysis}
\label{sec_2_complexity}

For comparison, a brief overview of the complexity of the discussed algorithms
is given in Table~\ref{table_complexity}. OMP/SD requires an OMP with $E$
iterations and one run of the SD with dimension (depth of the decoding tree)
$E$. SD-OMP requires to run the SD $E$ times, with dimensionality $i$ in the
$i$th iteration. Note, the complexity of these two algorithms depends only on
the sparsity but not on the dimension $L$ of the sparse vector.

In contrast, both pure-SD-based approaches depend on the dimensionality of the 
sparse vector $\ve{x}$ and are hence computationally infeasible for
high-dimensional problems. The SD with split matrix, proposed in
\cite{SDTeile,SDTeile_Verbessert}, needs to solve an $K$-dimensional problem
up to $|\A|^{L-K}$ times. The SD with enlarged matrix \cite{SD_augMatrix}
results in an $L$-dimensional problem, which, moreover, tends to be ill
conditioned.
\begin{table*}
\caption{\label{table_complexity}
	Complexity (number of iterations in OMP and dimensionality (depth of the
	decoidng tree) in SD) of the decoding approaches.
	Dimensions: $s \le E < K \ll L$.
}
\centerline{%
\def\arraystretch{1.2}\tabcolsep8pt
\begin{tabular}{l||c|c|c}
\hline
Algorithm	& Detection of
		& \#iter in OMP	& Dim.\ of SD
\\\hline\hline
OMP/SD		& $\ve{x}_\S$ after OMP
		& $E$
		& $E$ 
\\
SD-OMP		& $\ve{x}_\S$ within OMP
		& $E$
		& $1$, $\ldots$, $E$
\\
SD, split matrix \protect\cite{SDTeile,SDTeile_Verbessert}
		& $\ve{x}$
		& ---
		& $K$ (up to $|\A|^{L-K}$ times)
\\
SD, enlarged matrix \protect\cite{SD_augMatrix}
		& $\ve{x}$
		& ---
		& $L$
\\
\hline
\end{tabular}}
\end{table*}

\def\SER{\mathrm{SER}}
\def\Expt{\mathrm{E}}

\section{Numerical Results}
\label{sec_3}

\noindent
In this section, the performance of the proposed algorithms is evaluated in
terms of the symbol error rate $\SER = \Expt\{ \hat{x}_i \neq x_i \}$ by
numerical simulations. The measurement matrix $\ve{A}$ is obtained by
randomly selecting $K$ rows from a $L \times L$ unitary matrix and then
normalizing the columns to unit norm.

\subsubsection{Number of Iterations in OMP}
\label{sec_numAdd}

\begin{figure}[tb]
\psfrag{E}[c][b]{\small $E \ \longrightarrow$}
\psfrag{SER}[c][c]{\small $\SER \ \longrightarrow$}
\centerline{\includegraphics[width=.48\textwidth]{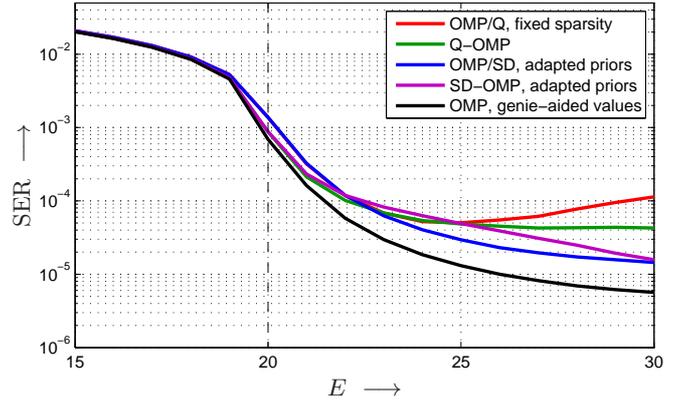}}
\caption{\label{fig_numAdd}
	SER of the proposed variants of OMP over the number $E$ of iterations.
	$L=256$, $K=128$, $s=20$, $\sigman2 \equiv 18\dB$.
}
\end{figure}

Fig.~\ref{fig_numAdd} shows the SER over the number of iterations, $E$, in the
OMP for $\sigman2 \equiv 18\dB$. The sparsity $s = 20$ is marked by a dashed
black line. In each case, additional iterations are rewarding.
If OMP/Q (red, symbol-wise quantization after OMP guaranteeing sparsity $s$)
is used, $E$ has to be selected carefully---choosing $E$ too large, the signal
estimation via the pseudoinverse fails and causes a degradation. Q-OMP%
\footnote{The thresholds of the ternary quantizer are optimized to $\pm 0.6$.
}
(green) does only reach similar performance but is much more tolerant to the
choice of $E$, which may be an advantage if the sparsity $s$ is not known
exactly.

Increasing $E$ the probability that the set $\S$ contains the correct support
increases (and tends to one as $E \to L$). In turn, the SD (OMP/SD, blue) is
able to most likely recover it. However, the gain comes at the cost of higher
computational complexity (larger dimensionality of SD).
OMP with embedded SD (SD-OMP, purple) shows a slightly better performance than
OMP/SD for $E \approx s$, but for $E > s$ OMP/SD outperforms SD-OMP et even
lower computational complexity. The problem in Q-OMP and SD-OMP is that the
residual $\ve{r}$ (Line~7 of Alg.~\ref{Algo_OMP}) is no longer orthogonal to
$\ve{\hat{x}}$, which affects the selection of the next support elements.

For reference, the SER when OMP finds the (enlarged) support set but perfect
(error-free) decisions of these symbols would be obtained at the final
quantization/sphere decoding step is shown (OMP, genie-aided values).
Via this curve it can be concluded that the main problem is to find a set
$\S$ which indeed contains the correct support. Using the SD the MIMO detection
problem is solved almost perfectly.

\subsubsection{Variants of OMP}
\label{sec_results_OMP}

\begin{figure}[tb]
\psfrag{EbNo}[c][b]{\small $10\log_{10}(\sigman2) \ [\mathrm{dB}] \
							\longrightarrow$}
\psfrag{SER}[c][c]{\small $\SER\ \longrightarrow$}
\centerline{\includegraphics[width=.48\textwidth]{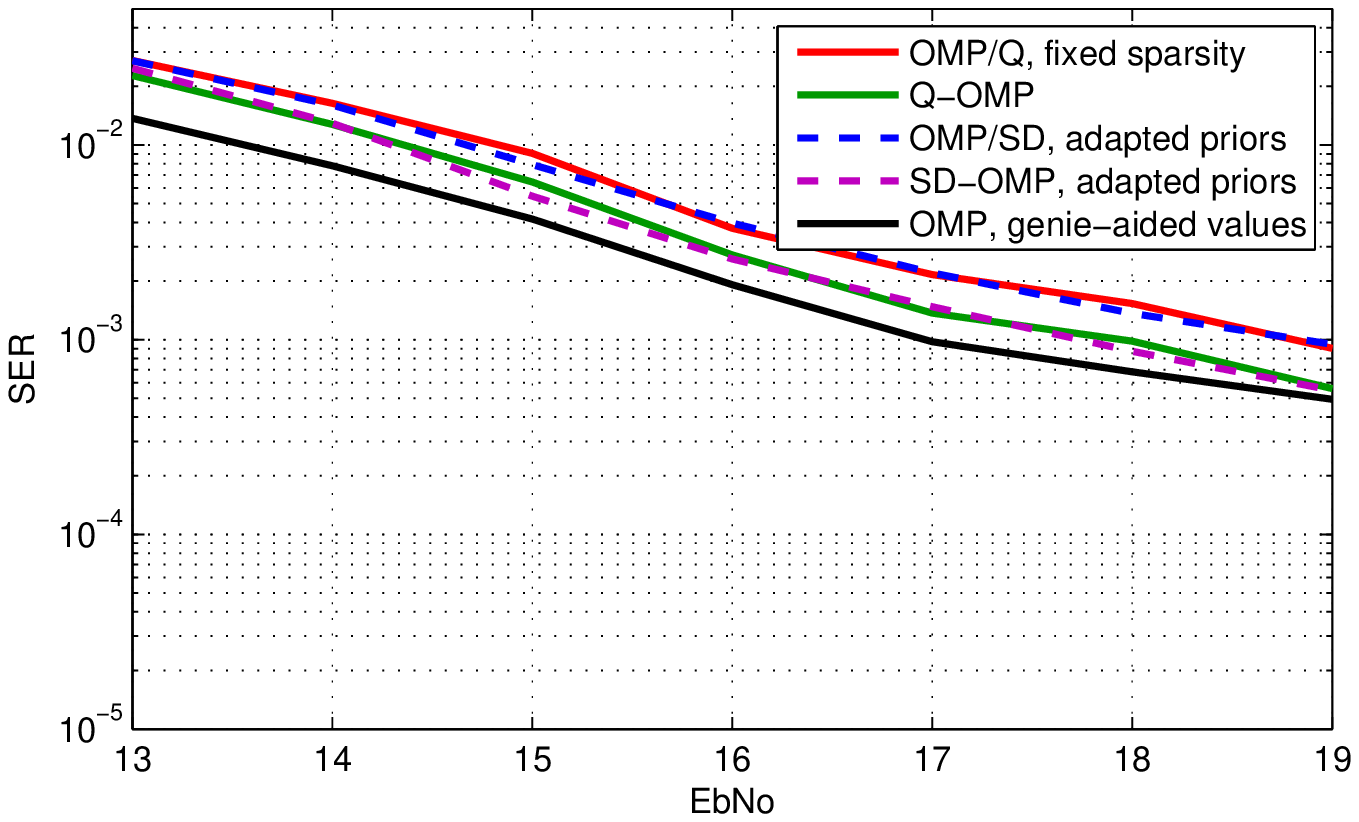}}
\caption{\label{fig_OMP_a0}
	SER of the proposed variants of OMP over the noise level $\sigman2$
	in dB.
	$L=256$, $K=128$, $s=20$. $E=20$.
}
\bigskip
\psfrag{EbNo}[c][b]{\small $10\log_{10}(\sigman2) \ [\mathrm{dB}] \
							\longrightarrow$}
\psfrag{SER}[c][c]{\small $\SER\ \longrightarrow$}
\centerline{\includegraphics[width=.48\textwidth]{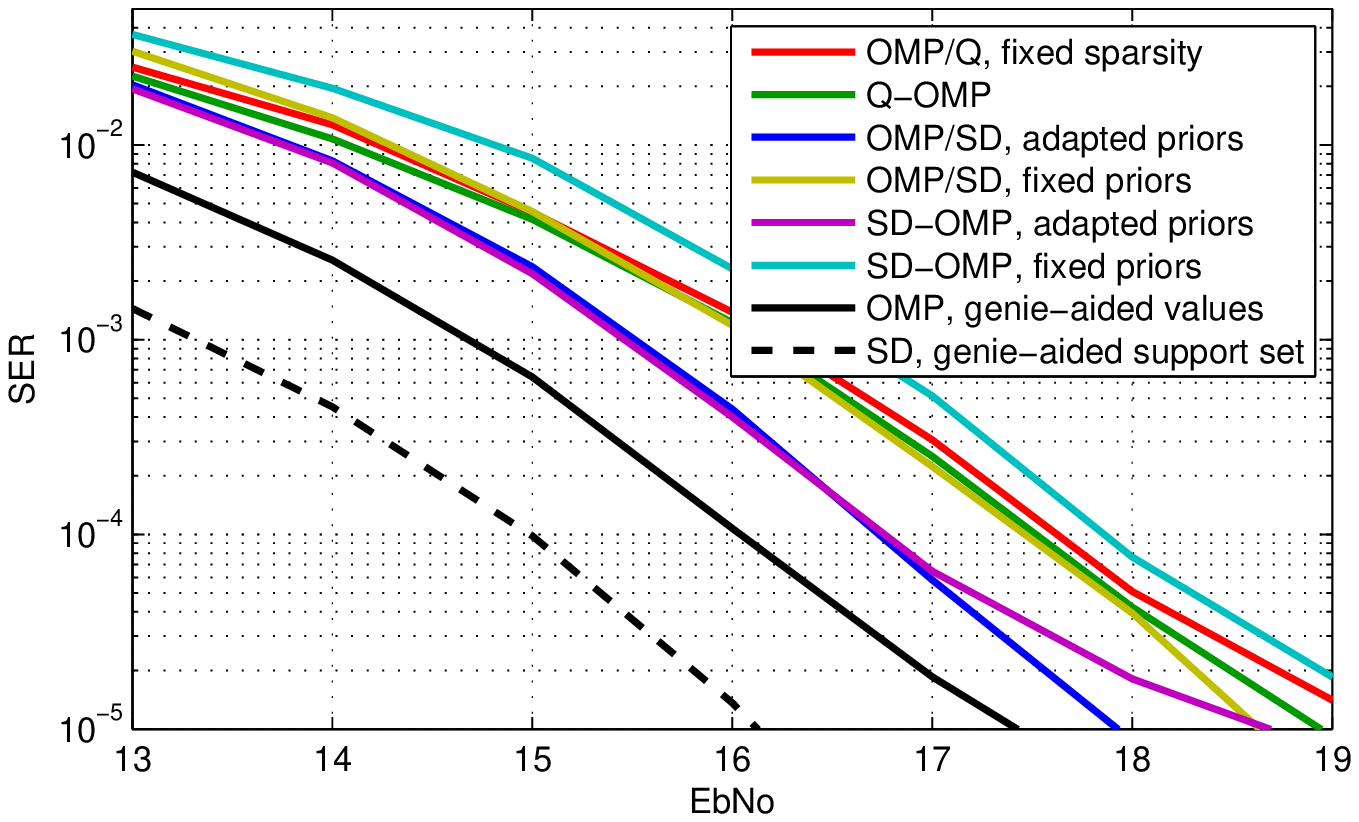}}
\caption{\label{fig_OMP_a10}
	SER of the proposed variants of OMP over the noise level $\sigman2$
	in dB.
	$L=256$, $K=128$, $s=20$. $E=24$ for QMP/Q; $E=30$ else.
}
\end{figure}

The above discussed variants of OMP are compared in Figs.~\ref{fig_OMP_a0} and
\ref{fig_OMP_a10}, where the SER is plotted over $\sigman2$ in dB.
Fig.~\ref{fig_OMP_a0} shows the results for the common approach $E = s = 20$,
while in Fig.~\ref{fig_OMP_a10} the number $E$ of iterations is optimized.
In view of Fig.~\ref{fig_numAdd} we choose $E = 24$ for OMP/Q and (to limit
complexity) $E = 30$ for all other approaches.

For $E = s$, a quantization embedded in the OMP gives slight gains at
negligible computational effort. Using the SD instead of scalar quantization
does not enable further gains but would only waste complexity. Once again,
as can be seen from the genie-aided reference curve, the problem is that the
OMP does not provide the correct support set.

Choosing $E > s$ the situation changes. Quantization embedded in the OMP does
not gain in performance compared to OMP with subsequent quantization. Here,
the additional iterations provide some tolerance of the algorithm to wrong
selections of support elements. If the sparsity is not known exactly, one can
benefit from the robustness of Q-OMP against additional iterations.

Using the SD for detection clearly outperforms symbol-wise quantization as long
as adapted a-priori probabilities are taken into account. From this
observation one can clearly conclude that by constraining the final detection
step, which is based on the enlarged set $\S$, to the correct sparsity $s$
(which is assumed to be known) significant gains are possible. Neither OMP/Q,
Q-QMP nor OMP/SD, SD-OMP with fixed priors guarantee the correct sparsity.

In summary, allowing the OMP to carry out some additional iterations, the
embedding of quantization or even the SD is not rewarding. The combination
of i) selecting an enlarged set of candidate positions for the support via OMP
and ii) detecting the discrete-valued symbols at these positions via SD is
a powerful and efficient approach. Noteworthy, the known sparsity should be
utilized and the SD has to be adapted to guarantee this fact.
Again, comparing the performance of proposed approaches with the genie-aided
curves clearly indicate the source of losses. If perfect decisions were
available for the positions, actually provided by OMP, only approximately
$0.5\dB$ could be gained (solid black curve). Conversely, if the correct
support set (plus random extra positions) was guaranteed to be included in
$\S$ and the SD worked on this set, much better performance would be possible
(dashed black).

\section{Conclusions}
\label{sec_4}

\noindent
In this paper, we have proposed and assessed approaches for the recovery of 
discrete-valued sparse signals. Combining CS algorithms, which essentially
serve to find a candidate set $\S$ which contains the correct support, and
subsequent MIMO detection schemes, in particular the sphere decoder, which
recover the discrete symbols, efficient low-complexity approaches are enabled.
Choosing the number of iterations of OMP large enough, an embedding of the MIMO
detection into the algorithm is not required. The adaptation of the decoding
metric in the SD, guaranteeing the desired/known sparsity, is crucial.

However, as the genie-aided reference curves show, the probability that the
correct support set is found should be increased. One way is to use the
CoSaMP instead of the OMP; almost everything shown for the OMP is equivalently
valid for the CoSaMP. Its known performance gains over OMP can be transferred
to the present situation of discrete sparse signals. An even more reliable
support set recovery, taking into account the finite nature of the symbols---%
in particular via reliabilities in the selection of the support positions---,
is still a field of current research.

\end{document}

%% file: xfig_csmimo.pstex_t
\begin{picture}(0,0)%
\includegraphics{xfig_csmimo.pstex}%
\end{picture}%
\setlength{\unitlength}{2901sp}%
\begingroup\makeatletter\ifx\SetFigFont\undefined%
\gdef\SetFigFont#1#2#3#4#5{%
  \reset@font\fontsize{#1}{#2pt}%
  \fontfamily{#3}\fontseries{#4}\fontshape{#5}%
  \selectfont}%
\fi\endgroup%
\begin{picture}(5468,637)(-111,-287)
\put(3691,179){\makebox(0,0)[b]{\smash{{\SetFigFont{8}{9.6}{\sfdefault}{\mddefault}{\updefault}{\color[rgb]{0.267,0.267,0.267}``sparsify''}%
}}}}
\put(1553,179){\makebox(0,0)[b]{\smash{{\SetFigFont{8}{9.6}{\sfdefault}{\mddefault}{\updefault}{\color[rgb]{0.267,0.267,0.267}``quantize''}%
}}}}
\put(2611, 74){\makebox(0,0)[b]{\smash{{\SetFigFont{8}{9.6}{\sfdefault}{\mddefault}{\updefault}{\color[rgb]{0.267,0.267,0.267}Discrete}%
}}}}
\put(2611,-151){\makebox(0,0)[b]{\smash{{\SetFigFont{8}{9.6}{\sfdefault}{\mddefault}{\updefault}{\color[rgb]{0.267,0.267,0.267}CS}%
}}}}
\put(4771, 74){\makebox(0,0)[b]{\smash{{\SetFigFont{8}{9.6}{\sfdefault}{\mddefault}{\updefault}{\color[rgb]{0.267,0.267,0.267}MIMO}%
}}}}
\put(4771,-151){\makebox(0,0)[b]{\smash{{\SetFigFont{8}{9.6}{\sfdefault}{\mddefault}{\updefault}{\color[rgb]{0.267,0.267,0.267}Schemes}%
}}}}
\put(451, 74){\makebox(0,0)[b]{\smash{{\SetFigFont{8}{9.6}{\sfdefault}{\mddefault}{\updefault}{\color[rgb]{0.267,0.267,0.267}Compressed}%
}}}}
\put(451,-151){\makebox(0,0)[b]{\smash{{\SetFigFont{8}{9.6}{\sfdefault}{\mddefault}{\updefault}{\color[rgb]{0.267,0.267,0.267}Sensing}%
}}}}
\end{picture}%